 \documentclass[acmlarge,screen]{acmart}

\usepackage{fontawesome}

\AtBeginDocument{%
  \providecommand\BibTeX{{%
    \normalfont B\kern-0.5em{\scshape i\kern-0.25em b}\kern-0.8em\TeX}}}

\setcopyright{acmlicensed}
\copyrightyear{2024}
\acmYear{2024}
\acmDOI{XXXXXXX.XXXXXXX}

\acmConference[ADVANCE '24]{International Workshop on ADVANCEs in ICT Infrastructures and Services}{February 26--29, 2024}{Hanoi, Vietnam}
\acmBooktitle{ADVANCE '24: Proceedings of the International Workshop on ADVANCEs in ICT Infrastructures and Services, February 26--29, 2024, Hanoi, Vietnam} 
\acmISBN{978-1-4503-XXXX-X/18/06}

\begin{document}

\title{Intelligent Data-Driven Architectural Features Orchestration for Network Slicing}

\author{Rodrigo Moreira}
\authornote{All authors contributed equally to this research.}
\email{joberto.martins@gmail.com}
\orcid{0000-0003-1310-9366}
\authornotemark[1]
\email{rodrigo@ufv.br}
\affiliation{%
  \institution{Universidade Federal de Viçosa (UFV)}
  \streetaddress{P.O. Box 1212}
  \city{Viçosa}
  \state{Minas Gerais}
  \country{Brazil}
  \postcode{}
}

\author{Flávio de Oliveira Silva}
\orcid{0000-0001-7051-7396}
\affiliation{%
  \institution{Universidade Federal de Uberlândia (UFU)}
  \streetaddress{Av. João Naves de Ávila, 2121 - CEP 38400-902}
  \city{Uberlândia}
  \country{Brazil}
}
\affiliation{%
  \institution{Universidade do Minho}
  \streetaddress{R. da Universidade, CP 4710-057}
  \city{Braga}
  \country{Portugal}
}

\author{Tereza Cristina Melo de Brito Carvalho}
\orcid{}
\affiliation{%
  \institution{Universidade de São Paulo (USP)}
  \city{São Paulo}
  \country{Brasil}
}

\author{Joberto S. B. Martins}
\orcid{0000-0003-1310-9366}
\affiliation{%
  \institution{Universidade Salvador (UNIFACS)}
  \streetaddress{Av. ACM 1133}
  \city{Salvador}
  \country{Brazil}}
\email{joberto.martins@ieee.org}

\renewcommand{\shortauthors}{Moreira and Martins}

\begin{abstract}
  Network slicing is a crucial enabler and a trend for the Next Generation Mobile Network (NGMN) and various other new systems like the Internet of Vehicles (IoV) and Industrial IoT (IIoT). Orchestration and machine learning are key elements with a crucial role in the network-slicing processes since the NS process needs to orchestrate resources and functionalities, and machine learning can potentially optimize the orchestration process. However, existing network-slicing architectures lack the ability to define intelligent approaches to orchestrate features and resources in the slicing process. This paper discusses machine learning-based orchestration of features and capabilities in network slicing architectures. Initially, the slice resource orchestration and allocation in the slicing planning, configuration, commissioning, and operation phases are analyzed. In sequence, we highlight the need for optimized architectural feature orchestration and recommend using ML-embed agents, federated learning intrinsic mechanisms for knowledge acquisition, and a data-driven approach embedded in the network slicing architecture. We further develop an architectural features orchestration case embedded in the SFI2 network slicing architecture. An attack prevention security mechanism is developed for the  SFI2 architecture using distributed embedded and cooperating ML agents. The case presented illustrates the architectural feature's orchestration process and benefits, highlighting its importance for the network slicing process. 

\end{abstract}

\ccsdesc[500]{Computing methodologies~Machine learning}
\ccsdesc[500]{Computing methodologies~Distributed artificial intelligence}
\ccsdesc[500]{Computing methodologies~Multi-agent systems}
\ccsdesc[500]{Computing methodologies~Cooperation and coordination}
\ccsdesc[500]{Networks~Programmable networks}
\ccsdesc[300]{Networks~Network management}
\ccsdesc[500]{Networks~Network design principles}
\ccsdesc[300]{Networks~Network dynamics}



\keywords{Network Slicing, Orchestration, Architectural Features, Intelligent Orchestration, ML-Native Slicing Architecture, Federated Learning, Data-Driven Slicing Architecture, SFI2, Security, ML-Native Security}

\begin{teaserfigure}
   \includegraphics[width=0.9\textwidth]{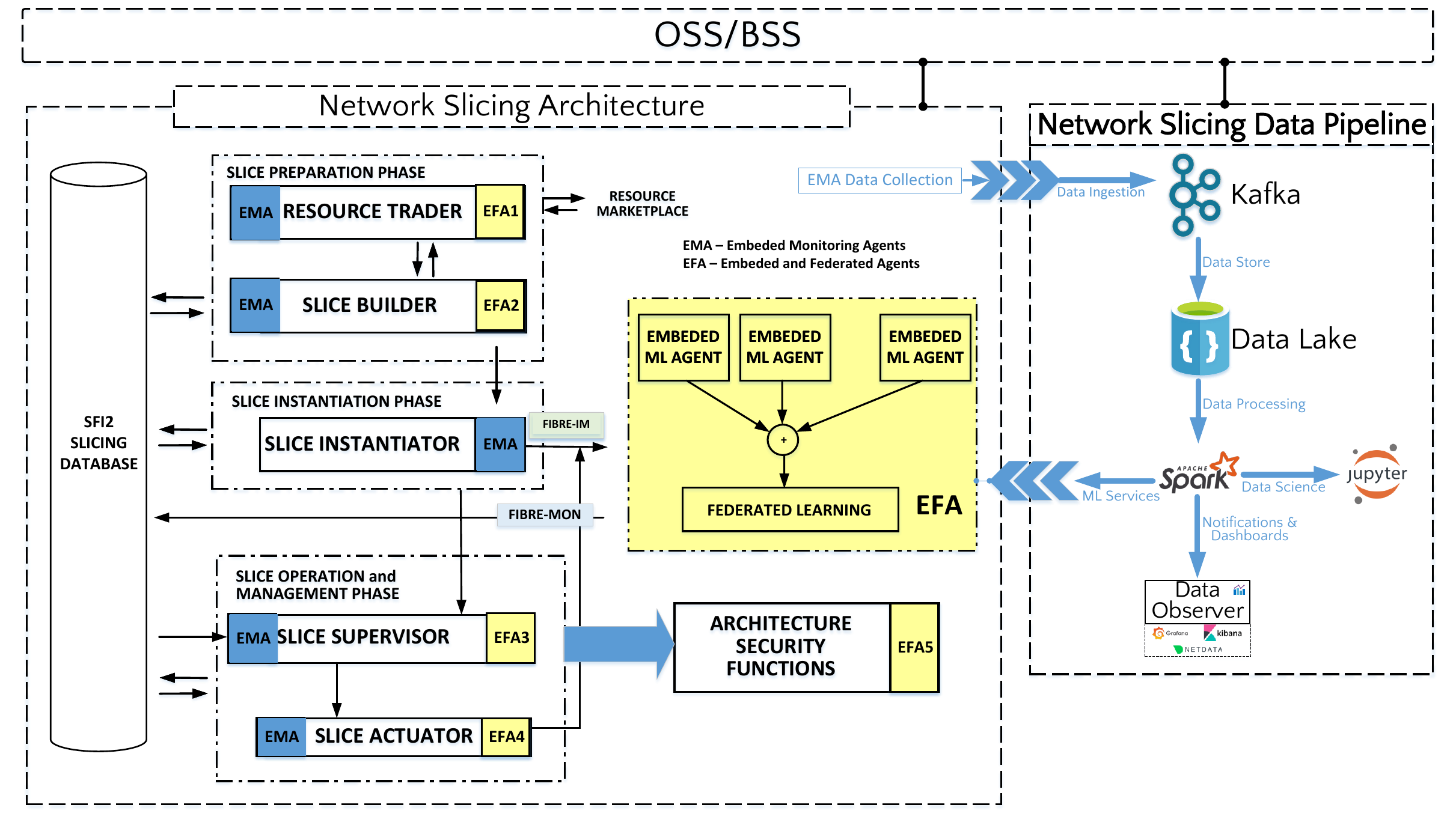}
  \caption{Intelligent Data-Driven Architectural Features Orchestration for Network Slicing}
  \Description{}
  \label{fig:teaser}
\end{teaserfigure}

\received{03 November 2023}
\received[revised]{30 December 2023}
\received[accepted]{10 January 2024}


\maketitle

\section{Introduction}\label{sec:intro}

Network slicing (NS) is a crucial enabler that supports virtual networks' planning, commissioning, configuration, operation, and management phases. Network slicing virtualizes physical resources like edge facilities, machines, communication links, switches, and radio access networks (RAN) while concomitantly allowing their customization and optimization \cite{subedi_network_2021} \cite{barakabitze_5g_2020}.

NS is adopted mainly in the next-generation mobile network (NGMN) (5G/6G) domain basically due to its inherent optimization capabilities that are necessary to accommodate the highly dynamic and variable requirements imposed by mobile users \cite{hong_6g_2022}. NS is also a trend in other domain areas such as Vehicular Networks \cite{waheed_comprehensive_2022}, experimental networks \cite{martins_enhancing_2023}, and industrial IoT \cite{wu_survey_2022}, among others, due to its capability to optimize and customize the network delivered for the user.

NS is an elaborated multi-phase process involving various architectural components. The existing network-slicing architectures like the ones proposed by 3GPP \cite{3gpp_3rd_2019}, IETF \cite{ietf_framework_2021}, ITU-T \cite{itu-t_framework_2012}, ETSI \cite{etsi_mobile_2015}, SFI2 project \cite{martins_enhancing_2023} and NECOS project \cite{clayman_necos_2021} aim to structure the overall slicing process. The NS process is structured by proposing architectural components, segmenting, and sequencing activities like preparation, commissioning, operation, and decommissioning. Distinct network-slicing architectures adopt approximately the same architectural components and use equivalent sequencing for the slicing process. Some variations exist in their component's structure and features and phase interrelations. Another aspect that varies among NS architectures is the architecture customization concerning the target user (ISPs, mobile users, experimental networks, among others). Finally, a common agreement point among standardization bodies and researchers is that artificial intelligence and machine learning are integral parts of the solution, focusing more specifically on optimizations\cite{phyu_machine_2023} \cite{nauman_artificial_2022}. 

However, although existing NS architectures have addressed NS phases and sequencing, solutions fall short of considering or recommending the orchestration strategy and approach among components, features, and resources. In this regard, machine learning utilization for NS optimization has to consider orchestration at various levels.

This paper addresses the issue of intelligent orchestration of features and resources towards a more efficient and robust network-slicing architecture. The proposed approach embeds intelligent agents in existing NS architecture components and phases to provide inherent services of various types and allow an improved architecture operation.

This paper is organized as follows. Section \ref{sec:intro} introduces the network slicing current scenario. Section \ref{sec:ArchOrch} presents the network slicing architectures and the architectural features orchestration in current NS architectures. Sections \ref{sec:PP} and \ref{sec:Data} present a set of architectural recommendations, followed by Section \ref{sec:Security} presenting an architectural features orchestration case for security. Finally, Section \ref{sec:Conclusion} closes the discussion with the final considerations.

\section{Network Slicing Architectures and Orchestration} \label{sec:ArchOrch}

Network slicing architectures share a common group of functional components and phases like the ones proposed by the 3GPP NS initiative \cite{3gpp_3rd_2019}. These common phases are:
\begin{itemize}
    \item Preparation Phase: - In the preparation phase, the slice request is received and interpreted, and the necessary resources are identified and localized in the resource market. In this phase, orchestration occurs in terms of selecting resources from multiple domains or multiple options available on a single domain.
    \item Commissioning Phase - This phase consists basically of making choices among the available resources aiming to configure the requested slice service. Orchestration at this phase occurs by making configuration choices that can potentially optimize resource allocation among deployed slices for a slice provider.
    \item Operation Phase - In the operation phase, the slice is already deployed and operational. Orchestration may occur in this phase, mainly due to dynamic user traffic patterns at different slice deployments.
    \item Decommissioning Phase- In this phase, the allocated slice resources from single or multi-domains are liberated.
\end{itemize}

As indicated, orchestration may occur at different steps of the network slicing process and inherently involves multiple components of a network slicing architecture.

\subsection{The SFI2 Network Slicing Reference Architecture} 

\begin{figure}
  \includegraphics[width=0.7\textwidth]{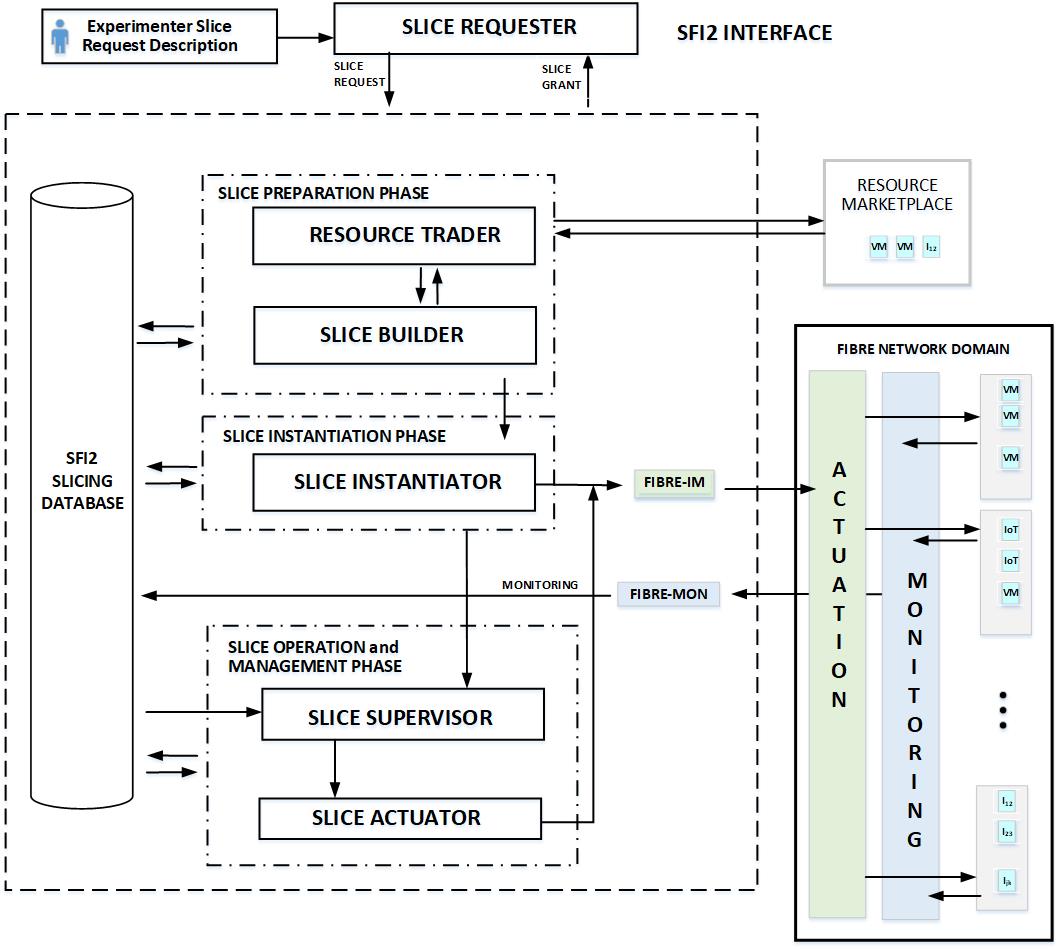}
  \caption{The SFI2 (Slicing Future Internet Infrastructures) Network Slicing Reference Architecture.}
  \label{fig:SFI2_Arch}
\end{figure}

The SFI2 project (Slicing Future Internet Infrastructures) defines a network slicing reference architecture, named SFI2 architecture, that aims to integrate experimental networks, incorporating architectural advances like ML-native optimizations, energy-efficient slicing, and slicing-tailored security functionalities \cite{martins_enhancing_2023} (Figura \ref{fig:SFI2_Arch}).

In Figure \ref{fig:SFI2_Arch}, the SFI2 architecture is deployed for the experimental FIBRE\footnote{FIBRE - Future Internet Brazilian Environment for Experimentation \cite{abelem_fit_2013}} domain and allows its users to create virtual slices across the 18 islands of the FIBRE network with virtual machines, IoT resources, and communications links. The SFI2 functionalities and operation are explored in sequence, aiming to identify the architectural features that are the object of discussion in this paper.

In the SFI2 FIBRE deployment, the list and description of resources that can be allocated to create user slices are available through the marketplace functionality. The FIBRE marketplace stores the list of resource descriptions that interact with SFI2 architecture during the preparation phase and can include some trading activities between the SFI2 FIBRE deployment and the resource provider, in this case, the FIBRE domain.

The slice builder builds the requested user slice considering the resources obtained from the marketplace and may optimize the utilization of these resources concerning the set of currently allocated resources used by the set of actively deployed slices.

The SFI2 slice instantiation deploys the configured slice in the FIBRE domain through a customized instantiation manager and sets up the required monitoring facilities for slice monitoring.

Finally, as the name suggests, the slice supervisor manages the slice operation, allowing slice reconfiguration resulting from user traffic changes, performance parameters tunning, SLA (Service Level Agreement) adjustments,  or from the need to reconfigure slices aiming the optimized use of resources by the slice provider (SFI2).

\subsection{The Architectural Features Concept for Network Slicing}

The architectural feature concept for network slicing can be understood as follows:
\begin{itemize}
    \item A set of objectives and characteristics defined for the deployed virtual slice and the network slicing architecture as a whole in a slicing process.
\end{itemize} 

To illustrate the concept, it follows a set of non-exhaustive architectural features considered for the scope of the discussion in this paper:

\begin{itemize}
    \item Resource selection in the preparation phase;
    \item Security capabilities for the slice or the architecture;
    \item Optimization of resources for the slices; and
    \item Optimization of resources for the network slicing provider.
\end{itemize}

As far as these architectural features are concerned, there is an inherent need for the orchestration among components and other elements in the architecture to achieve the required characteristics or to optimize resources. For example, orchestration for optimizing resources for a slice may consider service profile prediction mechanisms in the context of the slice itself, service profile prediction for the set of slices hosted in a provider, and the overall provider resource distribution among slices currently in use. When considering different types of attacks, an architecture security service should evaluate and consider a number of different architecture components since many of them are vulnerable to a single type of attack \cite{moreira_enhancing_2023}. In summary, architectural features setup requires a suitable orchestration mechanism in the network slicing architectures.

\subsection{Architectural Features Orchestration in Network Slicing - Current Scenario}\label{sec:NS_state_of_the_art}

Current alternatives for network-slicing architectures include architectures defined by standardization bodies and research projects. Table \ref{tab:summary} illustrates how some of the most relevant architectures inherently consider or not in their deployments a set of architectural features, including:
\begin{itemize}
    \item The capability to make multiple choices among available resources and optimize them at the preparation phase using the marketplace functionality;
    \item The capability to make multiple choices among available resources and optimize them at the preparation phase with multiple domains;
    \item The capability to orchestrate physical and logical resources towards optimization;
    \item The capability to orchestrate architectural components towards improved security capabilities at slice and architecture levels;
    \item The capability to orchestrate the composition of slice resources toward optimization; and
    \item The capability to orchestrate provider resources towards either architecture or provider resources optimization.
\end{itemize}

\begin{table}[ht!]
\centering
\caption{\textbf{Summary of architectural features orchestration in current network slicing architectures (NE - Not explicitly defined)}.}
\label{tab:summary}
\resizebox{\textwidth}{!}{%
\begin{tabular}{c|c|c|c|c|c|c|c}
\textbf{\textbf{ARCHITECTURAL FEATURE}}          & \textbf{\textbf{SFI2}} & \textbf{\textbf{3GPP}} & \textbf{ITU-T} & \textbf{ETSI} & \textbf{IETF} & \textbf{NECOS} & \textbf{NASOR} \\ \hline
\textbf{Marketplace}                             & \faCircle                    & \faCircleO                     & \faCircleO             & \faCircleO            & \faCircleO            & \faCircle            & \faCircleO             \\ \hline
\textbf{Multi-domain}                            & \faCircle                    & \faCircle                    & \faCircle            & \faCircle           & \faCircle           & \faCircle            & \faCircle            \\ \hline
\textbf{Physical/virtual Resource Orchestration} & \faCircle                    & NE                     & NE             & NE            & NE            & \faCircle            & \faCircle            \\ \hline
\textbf{Security Orchestration}                  & \faCircle                    & NE                     &  \faAdjust       & NE            & \faCircleO            & \faCircleO             & \faCircleO             \\ \hline
\textbf{Slice Resource Orchestration}            & \faCircle                    & \faCircle                    & \faCircle            & \faCircle           & \faCircle           & \faCircle            & \faCircle            \\ \hline
\textbf{Provider Resource Orchestration}         & \faCircle                    & \faCircleO                     & \faCircleO             & \faCircleO            & \faCircleO            & \faCircleO             & \faCircle            \\ \hline
\end{tabular}%
}
\end{table}

The network slicing surveys in \cite{wu_survey_2022} \cite{barakabitze_5g_2020} \cite{shen_ai-assisted_2020} \cite{donatti_survey_2023} and the network slicing architectures presented in \cite{3gpp_3rd_2019}, \cite{clayman_necos_2021}, \cite{etsi_mobile_2015}, \cite{ietf_framework_2021}, \cite{itu-t_framework_2012}, \cite{martins_enhancing_2023}, and \cite{moreira2021} further detail the existing architectural features orchestration for NS. In Table \ref{tab:summary}, security orchestration refers to the ability to orchestrate security functionalities not only for slices but also for the architecture that provides these slices. Slice orchestration refers to the ability to orchestrate available resources in slice deployment, whereas provider orchestration refers to the capability to orchestrate resources among deployed slices in an NS architecture. We use the symbols \faCircle, \faCircleO, and \faAdjust to represent compliance with the feature, noncompliance, and partial compliance, respectively.

\section{Architectural Features Orchestration for Network Slicing Architectures - Position Points and Recommendations} \label{sec:PP}

Network slicing architectures and derived systems must be dynamic and efficient regarding resource orchestration and allocation. To achieve such characteristics, architectural features like the ones indicated in Table \ref{tab:summary} must be in place, orchestrating the available resources.

\begin{figure}[H]
  \includegraphics[width=0.8\textwidth]{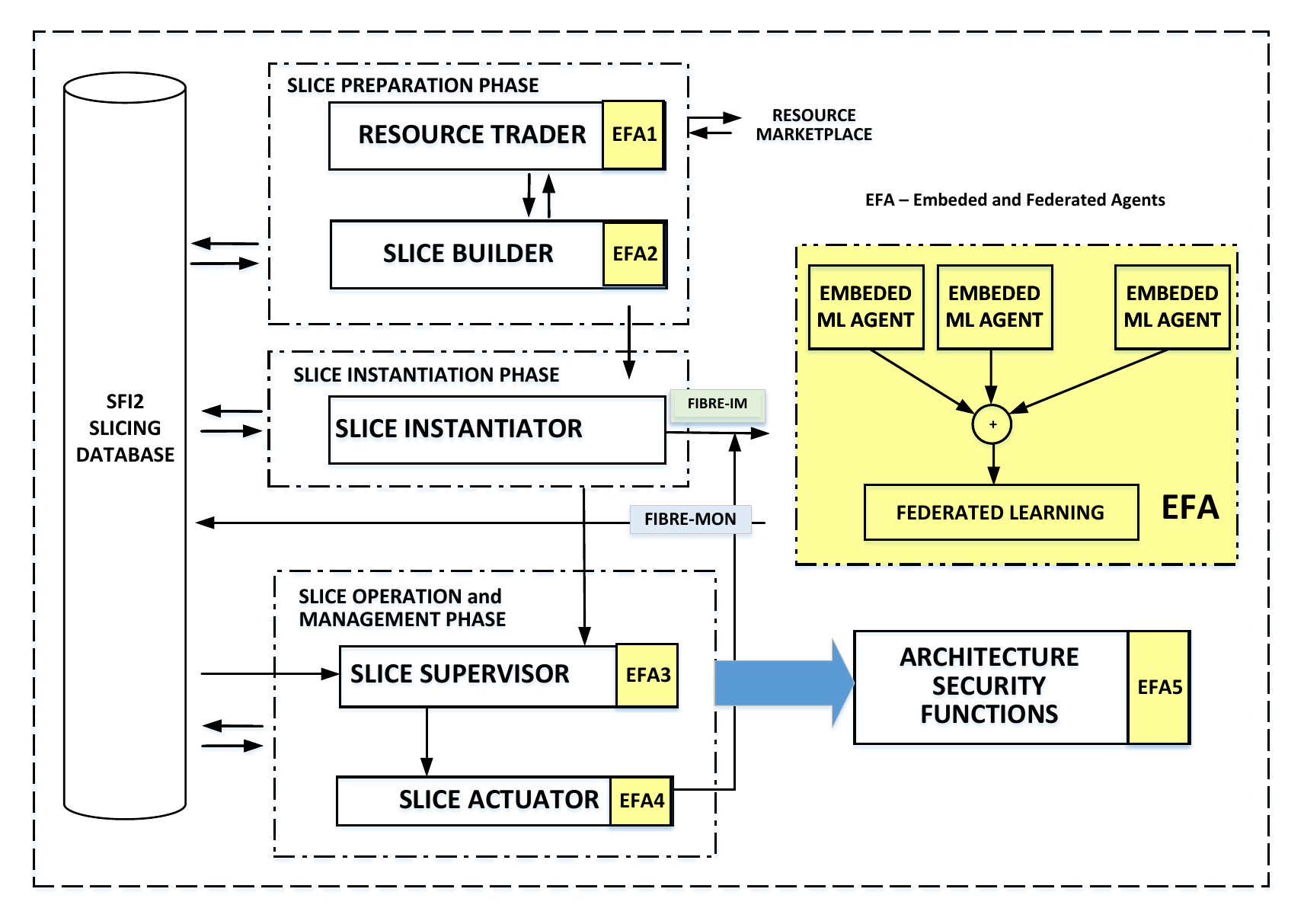}
  \caption{ML-Native Agents and Federated Learning Recommendations for Architectural Features Orchestration.}
  \label{fig:EFA}
\end{figure}

In this regard, we propose the following recommendations for network-slicing architectures that will allow  optimization and support architectural features deployment for slice providers and users (Figure \ref{fig:EFA}): \\

\begin{minipage}{\linewidth}
    \begin{itemize}
        \item Machine learning agents as an embedded and intrinsic functional component;
        \item Federated learning as a network-slicing architectural intelligence capability; and
        \item Data-driven methods to manage network slicing services.
    \end{itemize}
\end{minipage}
\\

Having machine learning agents as an embedded and intrinsic functional component means, in other words, including ML-native functionalities in the architecture. This is achieved, for instance, in the SFI2 reference architecture by having the basic phases (preparation, trading, building, operation, supervision, and security) of the network slicing process assisted by ML agents (Figure \ref{fig:EFA}).

However, embedding machine learning agents in a network-slicing architecture solves part of the objective to include a problem-solving intelligence we want to address. In effect, the different slicing phases in the network-slicing process must interact through the specific orchestration solution adopted in the architecture or corresponding deployment. In technical terms, this means that different ML agents in distinct architecture components have models that should be integrated in the best possible way. To attain this objective, we suggest that federated learning should be used to weigh among agents and arrive at a kind of weighted model for the specific orchestration in place.

For example, providing clean and energy-efficient slices across multi-domain resource providers requires the orchestration between distinct energy-efficient algorithms and approaches by providers that should be weighted to the type of clean or not-clean energy they use. In summary, information and knowledge are distributed and, as such, should be used considering these characteristics.

\section{Data-driven Methods Recommendation for Network Slicing}\label{sec:Data}

\begin{figure}[H]
  \includegraphics[width=0.9\textwidth]{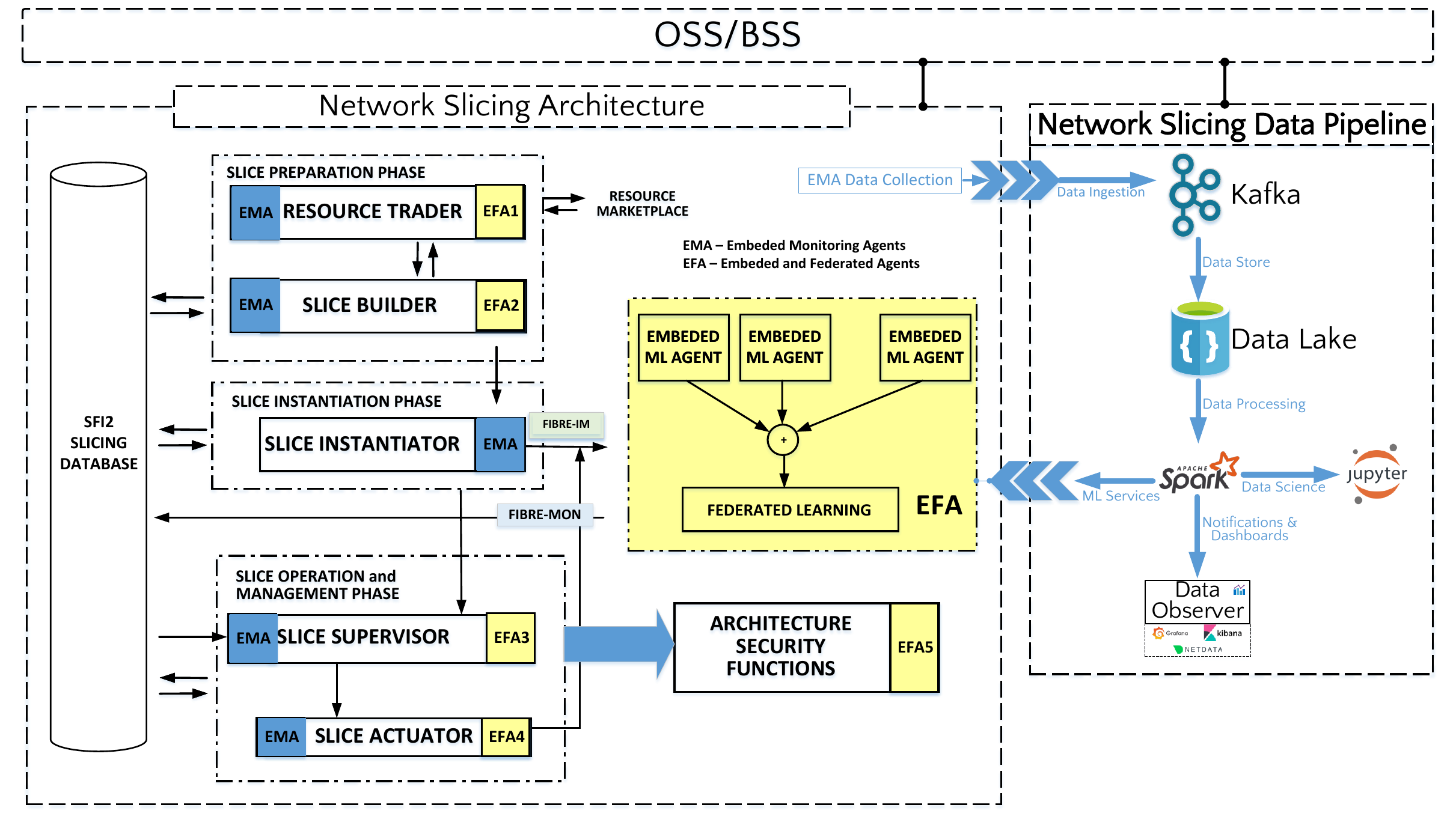}
  \caption{Architectural Data-Driven Recommendation for Network Slicing.}
  \label{fig:data_pipelining}
\end{figure}

Data management is a critical element in various areas, including the network architecture. Network slicing generates a large volume of data during different orchestration phases, and the service runtime on top of tailored resources is the primary data source. Network providers can utilize data-driven methods to manage their services and improve security and intelligent network-slicing services. With a vast amount of data available, modern network architectures can offer numerous insights and customized services. To achieve this, we propose a data pipeline, as shown in Figure~\ref{fig:data_pipelining}, that includes an Embedded Monitoring Agent (EMA) to collect data and facilitate data ingestion. Services such as Kafka handle data streams for batch and data processing, whereas later services such as Spark distribute streamlined data processing, enabling real-time analytics and seamless scalability. This approach shifts data management practices and fosters agile decision-making in network markets. In addition, some processing can feed EFAs or provide a data science playground for researchers through the Jupyter environment. Ultimately, the network-slicing user or management can gain insights and effectively monitor network slices.

Granular insights facilitate a novel approach to managing network resources and sharing legacy facilities via the Operations Support System and Business Support System (OSS/BSS). In addition, the data generated by both the applications running through network slicing and the operation of the network architecture can feed into cognitive methods to make communication seamless and intelligent. Many challenges concerning security, data management, and heterogeneous data sources must be considered for next-generation network-slicing methods.

\section{ML-native Security - An Architectural Feature Orchestration Case}\label{sec:Security}

The SFI2 slicing architecture is an edge-cutting approach that deploys intelligent, energy-efficient network slices while guaranteeing security at both the operational and service levels. Operational security is related to the safety of the architectural building block, whereas service refers to the security of the slicing service. Here, we highlight some of the key points of the architecture, particularly the mechanisms that enable the deployment of intelligent and secure network slices~\cite{martins_enhancing_2023}. The rationale behind our architecture inaugurates the native distributed machine-learning mechanism embedded into architecture building blocks. 

We devised a Machine Learning Agent (ML-Agent) mechanism, which has a dual responsibility: to perceive and act in the environment, the partner iterates over the data collected from the archive to train ML models in a distributed manner and report model weights to the main model in SFI2 AI Management. The mechanism of action in the intelligent network slicing process refers to the interaction between the ML-agent and network slicing functional blocks (Slice Builder, Supervisor, and Actuator in Fig.~\ref{fig:EFA}). In this way, for each life cycle of the network slice, the process can rely on machine learning models to make decisions such as resource allocation, provisioning, and Quality of Experience or Service (QoE/QoS) forecasting.

\begin{figure*}
  \includegraphics[width=0.8\textwidth]{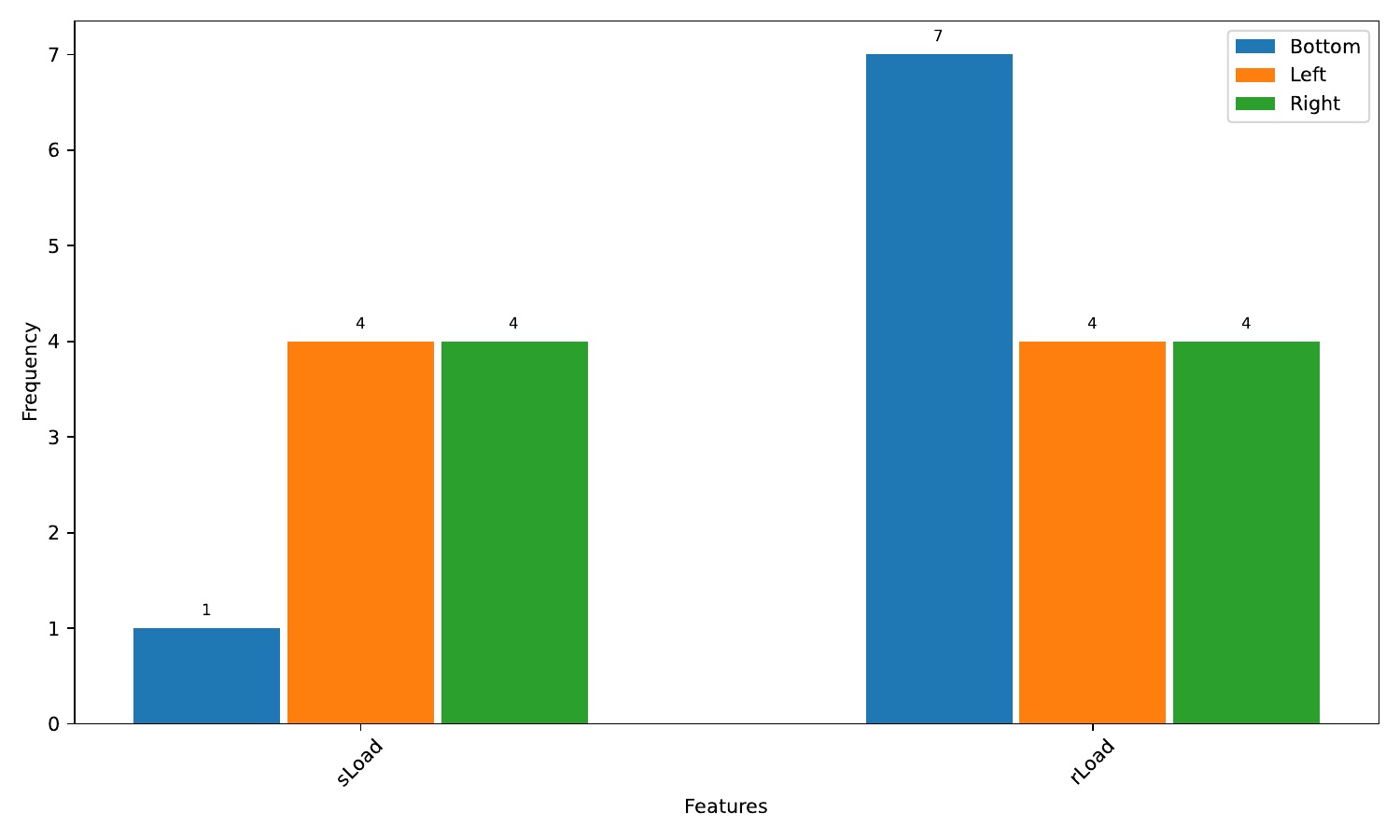}
  \caption{Frequency of the most important features in whole Westermo dataset.}
  \label{fig:feature_importante}
\end{figure*}

We examined three datasets simultaneously and applied feature engineering to determine the most important features for data clustering. We applied the elbow method~\cite{bholowalia_ebk-means_2014} to determine a suitable number for clustering our data using the k-means algorithm, which revealed eight (8) clusters. In the feature engineering scheme shown in Figure ~\ref{fig:feature_importante}, we noticed that \textit{rLoad} and \textit{sLoad} were the most important features that best clustered most of the data among the three datasets. The comparison illustrates the varying frequencies of essential features among datasets, implying potential distinctions and similarities within the data.

Knowing the most relevant features identified by the k-means clustering algorithm provides valuable insights for further analysis and model refinement. Regarding slicing architectures, feature orchestration is important once service-level agreements and threat systems can operate earlier, avoiding slices and architecture operation outages. Although these features may not exhibit strong linear correlations, their significance in cluster formation implies they carry essential information for distinguishing distinct groups within the dataset. For further investigations, we believe integrating data pipelining into network-slicing architectures will shift slicing management and orchestration in future network-slicing architectures.

Principal Component Analysis (PCA) was used to identify the most relevant features for each cluster. PCA is a dimensionality reduction technique that transforms data into a new coordinate space, where the axes are called principal components. The first principal component explains most of the data variance and so on. PCA allows us to obtain the coefficients of the principal components that indicate the weight of each original feature in forming the new axes. The feature with the highest absolute coefficient for each principal component is the most relevant for that axis.

Second, we showcase the architecture feature to handle training on different blocks of the SFI2 architecture using a Westermo dataset in our evaluation~\cite{Strandberg2023}. This dataset refers to 90 min of packet industrial network slicing, including harmless SSH, bad SSH, misconfigured IP addresses, duplicated IP addresses, port scans, and man-in-the-middle attacks. The context of the dataset refers to an industrial network, and three collections of PCAPs (Packet Capture) probes were distributed along the topology. Based on this dataset, we imported it into the architecture to validate the distributed training and prediction mechanism. We simultaneously considered and trained three datasets, implying a Non-Informally, Identically Distributed (non-IID) configuration. We employed traditional machine learning algorithms to assess the prediction suitability of the ML-Agent among the architectural block operations.

\begin{figure*}
  \includegraphics[width=0.7\textwidth]{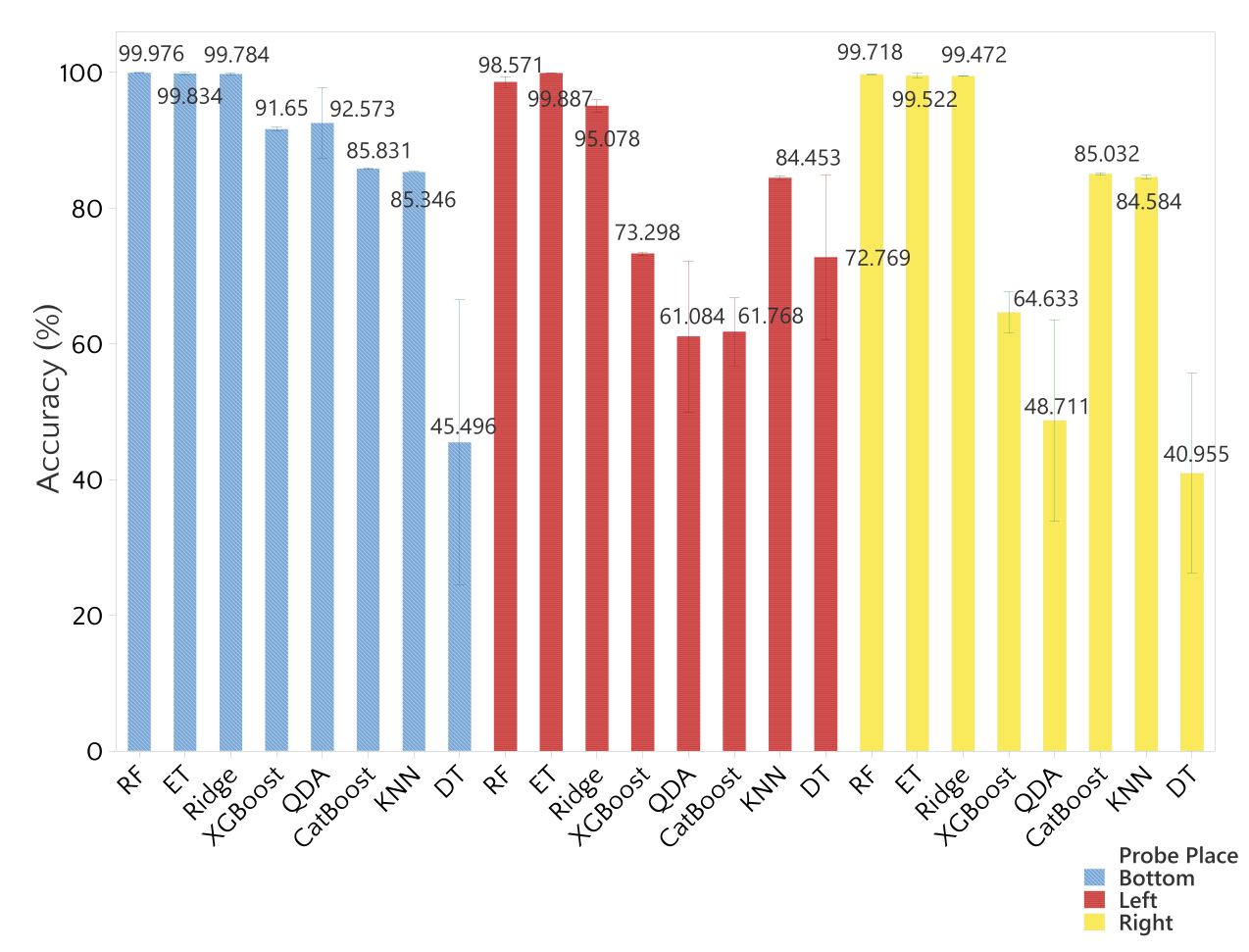}
  \caption{Localized Test Accuracy (\%)  achieved by three different ML-Agents in the SFI2 Slicing Architecture.}
  \label{fig:basic_algorithms_results}
\end{figure*}

We idealized an experiment considering machine learning algorithms to choose the best one that fits the network data empirically. Hence, our considered network topology has different equipment, a network manufacturing traffic simulator, and three packet probes at different network locations: bottom, left, and right. We empirically evaluated the following algorithms: extra threats classifier (ET), Random Forest Classifier (RF), Ridge, Quadratic Discriminant Analysis (QDA), Extreme Gradient Boost (XGBoost), CatBoost Classifier, K-Neighbors Classifier (KNN), and Decision Tree (DT) \cite{sarker_machine_2021} \cite{chaudhary_improved_2016} \cite{zhang_introduction_2016} \cite{song_decision_2015}. Each ML agent is placed over the network to evaluate the algorithms.

In summary, our experiments considered distributed training over three different ML-Agents for prediction in different architectural layers. We measured the convergence accuracy of the server model using local training and testing. Hence, we summarized our results concerning accuracy and loss over epochs according to Figure~\ref{fig:basic_algorithms_results}, where, despite the challenging training scenario with a non-IID dataset, the SFI2 AI Management block handles a model with the prediction of harmless packets with a lower error and 95\% confidence. Using locally trained models, the ML-Agent achieved higher accuracies of approximately 99\% for many algorithms. We conducted our experiments by considering the stratified KFold with ten (10) folds for each algorithm. This experiment validated the safety features of our architecture while exploring the ML-native slicing architecture, unlike the QDA and XGBoost algorithms, which performed well in some places of topology, but not at all.

\section{Final Considerations} \label{sec:Conclusion}

Dynamic and optimized orchestration of resources is the main driver of network slicing, which allows its adequacy and, at least in part, justifies its trend in areas such as the next-generation mobile network (5G/6G), IoV, and IIoT, where user requirements are highly stringent and heterogeneous services are required. This study proposes, highlights, and demonstrates that network-slicing architectures should incorporate ML-native agents in their structure, adopt a distributed learning strategy for acquiring knowledge, and incorporate a data-driven method to manage network-slicing services.

A \textit{security service} is an architectural feature demonstrated in the SFI2 architecture, in which embedded security agents use federated learning to acquire cooperative knowledge using a data-driven approach to provide intrusion detection. The architectural deployment for the SFI2 architecture can be replicated in other network slicing architectures by adapting the proposed approach and method demonstrated for the slicing phases and components in the target architecture. For future directions and research agenda, we believe that the coexistence of security, monitoring, and intelligent agents embedded in architectural services as daemons will be crucial for future generations of network-slicing architectures and services.  In addition, we guess that the hybrid and collaborative use of supervised and unsupervised learning paradigms is essential for discovering knowledge and generating decision-making insights for near-real-time network orchestrators and managers.

\begin{acks}

The authors thank the FAPESP MCTIC/CGI cooperation agreement under the thematic research project 2018/23097-3 - Slicing Future Internet Infrastructures (SFI2),  Brazilian National Council for Scientific and Technological Development (CNPq), grant \# 421944/2021-8, and the ANIMA INSTITUTE for scholarship support.

\end{acks}

\bibliographystyle{ACM-Reference-Format}
\bibliography{ADVANCE}

\appendix

\section{Research Methods}

\subsection{Part One}

The research method used in this paper is exploratory and quantitative. An exploratory method is achieved by analyzing, discussing, and recommending intelligent architectural feature orchestration for network slicing architectures. The quantitative method is achieved by simulating a case for embedding security in the SFI2 architecture.

\subsection{Part Two}

The security architectural feature orchestration case simulates three (3) ML agents deployed in distinct SFI2 architecture functional blocks (phases) with perceive and acting capabilities. The agents are trained using the extra threats classifier (ET), Random Forest Classifier (RF), Ridge, Quadratic Discriminant Analysis (QDA), Extreme Gradient Boost (XGBoost), CatBoost Classifier, K-Neighbors Classifier (KNN), and Decision Tree (DT) ML-classification methods with the Westermo dataset over relevant features detected. The federated learning average approach is used to measure convergence accuracy for the distributed training.

\section{Online Resources}

The security architectural feature case experimental dataset and code are available on GitHub: \url{https://github.com/romoreira/ADVANCE-IntrusionDetectionSystem}

\end{document}